\documentclass[letter]{aa}  
\usepackage{natbib}
\usepackage{graphicx}
\usepackage{times}
\usepackage{amsmath}
\usepackage{txfonts}
\newcommand{\msun}{{\,\rm M}_{\odot}}

\newcommand{\nht}{\ifmmode {{\rm NH}_3} \else {NH{\bas 3}} \fi}
\newcommand{\tco}{\ifmmode {^{13}{\rm CO}} \else {$^{13}{\rm CO}$}\fi}
\newcommand{\dco}{\ifmmode {^{12}{\rm CO}} \else {$^{12}{\rm CO}$}\fi}
\newcommand{\cdo}{\ifmmode {{\rm C}^{18}{\rm O}} \else {${\rm C}^{18}{\rm
O}$}\fi}

\newcommand{\htco}{\ifmmode {{\rm H}^{13}{\rm CO}^{+} } \else {${\rm H}^{13}
{\rm CO}^{+}$ }\fi}
\newcommand{\hco}{\ifmmode {{\rm H}^{12}{\rm CO}^{+} } \else {${\rm H}^{12}
{\rm CO}^{+}$ }\fi}
\newcommand{\ndhp}{\ifmmode {{\rm N}_{2}{\rm H}^{+} } \else {${\rm N}_{2}
{\rm H}^{+}$ }\fi}
\newcommand{\juz}{\ifmmode {{\rm J}=1\rightarrow 0} \else
{J=1$\rightarrow$0}\fi}
\newcommand{\jdu}{\ifmmode {{\rm J}=2\rightarrow 1} \else
{J=2$\rightarrow$1}\fi}
\newcommand{\jtd}{\ifmmode {{\rm J}=3\!\rightarrow\!2} \else
{${\rm J}=3\!\rightarrow\!2$} \fi}
\newcommand{\jcq}{\ifmmode {{\rm J}=5\!\rightarrow\!4} \else
{${\rm J}=5\!\rightarrow\!4$} \fi}
\newcommand{\as}{\ifmmode {^{\scriptscriptstyle\prime\prime}}
        \else $^{\scriptscriptstyle\prime\prime}$\fi}
\newcommand{\am}{\ifmmode {^{\scriptscriptstyle\prime}}
        \else $^{\scriptscriptstyle\prime}$\fi}
\newcommand{\TableRing}{
\begin{table}
\begin{center}
\begin{tabular}{lcc}
\multicolumn{3}{c}{Ring observational parameters} \\
 \hline \hline
                  & This work       & Guilloteau et al.\ 1999 \\
Frequency (GHz)   & 267             & 220  \\
Ang. resolution($''$)  & $0.45\times0.25$ & $0.88\times0.56$ \\
 \hline
Flux (mJy)        & $760 \pm 10$    & $544  \pm 3$ \\
Inner radius ($''$) & $1.38 \pm 0.01$ & $1.27 \pm 0.01$ \\
Outer radius ($''$) & $1.84 \pm 0.01$ & $1.90 \pm 0.01$ \\
 \hline
 \end{tabular}
\caption{Parameters of a geometrical ring fitted to the $uv$ data,
and comparison with the results obtained with previous data.
\label{tab:ring} }
\end{center}
\end{table}
}
\newcommand{\TableModel}{
\begin{table}
\begin{center}
\begin{tabular}{lc}
\multicolumn{2}{c}{Ring physical parameters} \\
 \hline \hline
Inclination ($^\circ$)      &   $ 37 \pm 1$  \\
Position angle ($^\circ$)   &   $  7 \pm 1$ \\
\hline
Inner radius (AU)           &   $200\pm2$ \\
Outer radius (AU)           &   $250\pm2$ \\
Edge width   (AU)           &   $<10$ \\
\hline
Temperature (K)             &   [35] \\
Temperature exponent        &   [0] \\
\hline
Surface density (cm$^{-2}$) &   2.2$\pm0.05 \times$10$^{24}$ \\
Surface density (g.cm$^{-2}$) &   9.6$\pm0.2$ \\
\hline
Dust emissivity $\kappa$ (cm$^{-2}$/g)      &   [0.012]      \\
\hline
 \end{tabular}
\caption{Fitted parameters of the GG~Tau circumbinary ring. Error bars
are 1 $\sigma$ formal errors, not taking into account phase noise nor
flux calibration uncertainty. Square brackets indicate fixed
parameters.\label{tab:model} }
\end{center}
\end{table}
}
\newcommand{\FigVisi}{%
  \begin{figure*}
    \centering %
    \includegraphics[angle=270,width=\textwidth]{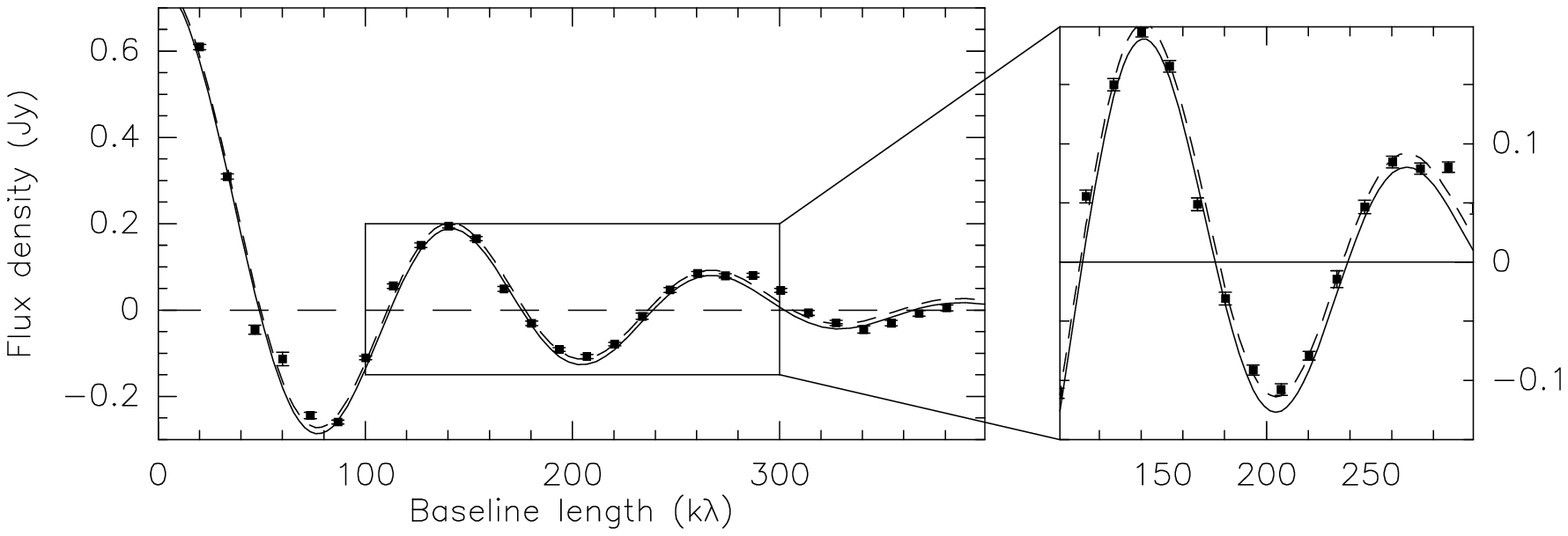}
    \caption{Real part of the visibility $vs$ baseline lengths. The
    points and error bars represent the measured values, and the
    curves are the best fits of a geometrical ring (plain curve; see
    Table~\ref{tab:ring}) and of a ring plus a point source
    (dashed). The visibilities have been corrected from the disk
    orientation ($PA=7^{\circ}$) and inclination ($i=37^{\circ}$) by 
    compressing the V value by $\cos(i)$ prior to circular averaging.}
  \label{fig:visib}
  \end{figure*} }
\newcommand{\FigMaps}{%
  \begin{figure*}
    \centering %
    \includegraphics[angle=270,width=\textwidth]{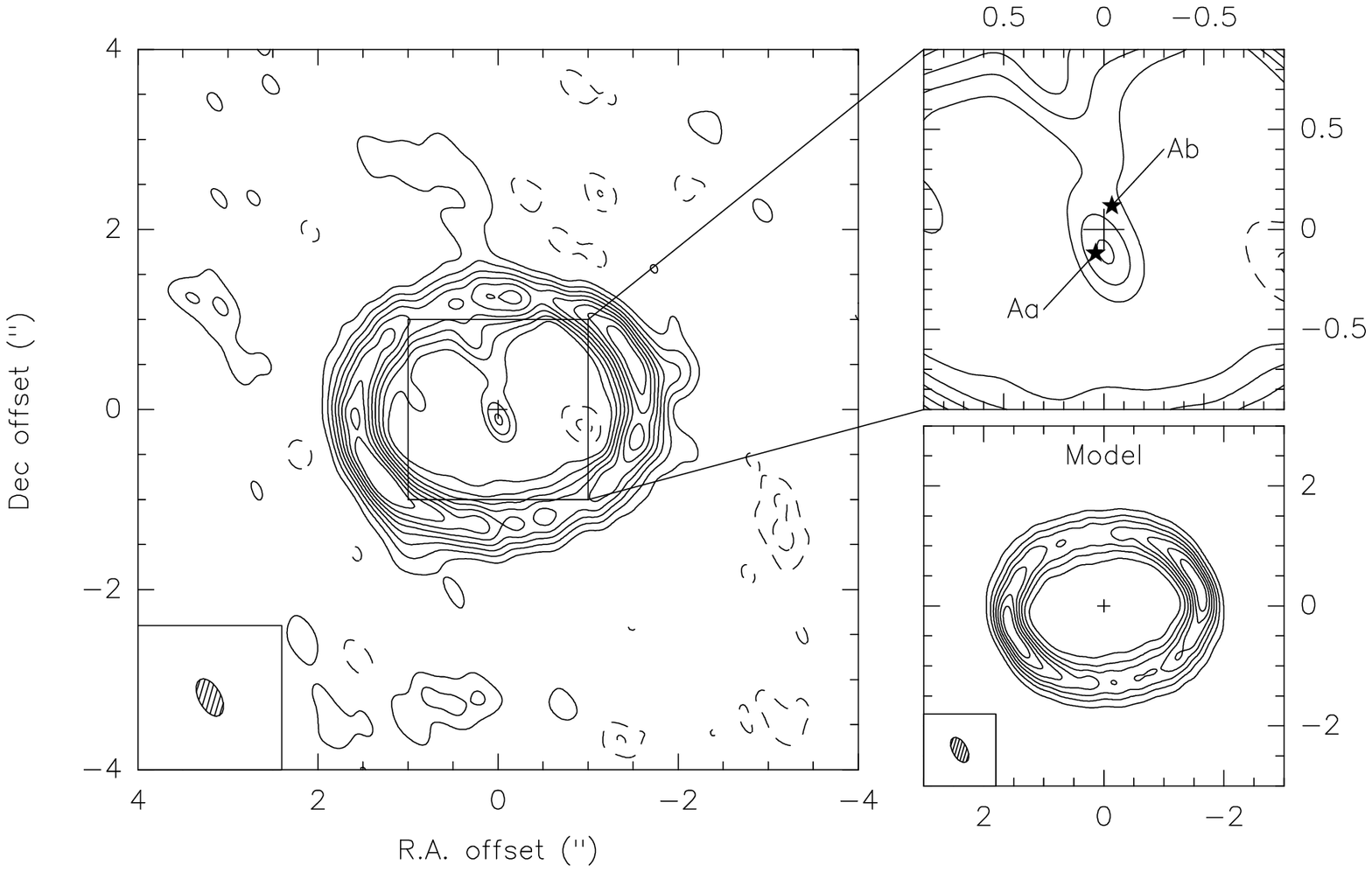} 
    \caption{{\bf Left}: 267~GHz continuum image of GG\,Tau. The
    angular resolution is $0.45\as \times 0.25\as$ at PA $19^\circ$,
    and the contour spacing is 3.1~mJy/beam (0.47~K, about
    2~$\sigma$). The map is centered at $\alpha =04:32:30.356$,
    $\delta=17:31:40.48$ (cross, J2000). The GG\,Tau\,B binary is
    located $\sim$10$''$ towards the south, outside the plotted
    area. {\bf Top right}: zoom of the inner 1$''$ and position of the
    GG Tau A binary stars as of 2008. {\bf Bottom right}: best
    physical ring model (see Table \ref{tab:model} and text). Angular
    resolution and contour spacing are as in the observations.}
  \label{fig:maps}
  \end{figure*}}
\begin{document}
%
%---------------------------------------------------------------------------
%
\title{High resolution imaging of the GG Tau system at 267 GHz
\thanks{Based on observations carried out with the IRAM Plateau de
Bure Interferometer. IRAM is supported by INSU/CNRS (France), MPG
(Germany) and IGN (Spain).}}

\author{Vincent Pi\'etu \inst{1} \and Fr\'ed\'eric Gueth \inst{1} \and
Pierre Hily-Blant\inst{1,2} \and Karl-Friedrich Schuster \inst{1} \and
Jer\^ome Pety \inst{1}}
\offprints{Vincent Pi\'etu, \email{pietu@iram.fr}}
\institute{IRAM, 300 rue de la piscine, 38406 Saint Martin d'H\`eres,
France \and LAOG,Universit\'e Joseph Fourier, CNRS, 38041 Grenoble,
France}
\date{Received 2 September 2010/ Accepted 9 December 2010}
%
%---------------------------------------------------------------------------
%
\abstract
% Context
{Studying circumbinary disks is critical to understanding the formation
mechanisms of binary stars. While optical or mid-infrared images
reveal the scattered emission, millimeter observations provide direct
measurements of the dust thermal emission.}
% Aims
{We study the properties of the circumbinary disk around the
well-known, multiple young stellar object GG\,Tau with the highest
possible sensitivity and spatial resolution.}
% Methods
{We mapped the continuum emission of GG\,Tau at 267~GHz using the IRAM
Plateau de Bure interferometer equipped with upgraded receivers and LO
systems. An angular resolution of $0.45''\times0.25''$ was
achieved, corresponding to a linear resolution of $65\times35$ AU.}
% Results
{The GG\,Tau\,A circumbinary disk is observed as an extremely clearly
defined narrow ring. The width of the ring is not resolved. Emission
from the central binary is detected and clearly separated from the
ring: it coincides with the GG\,Tau\,Aa position and may therefore
trace a circumstellar disk around this star. The mass ratio of
the circumbinary to circumprimary material is $\sim$80.}
% Conclusions 
{} 
%
%---------------------------------------------------------------------------
%
\titlerunning{0.3'' imaging of the GG Tau disk} 
\authorrunning{Pi\'etu et al.}  
\maketitle
%
%---------------------------------------------------------------------------
%
\section{Introduction}

GG\,Tau is a well-known quadruple system located in the Taurus
star-forming region. It is composed of two binaries, the northern one
(GG Tau A) having been the subject of considerable attention during
the past decade.  This system, classified as a classical T\,Tauri
star, is a close binary with a separation of $\sim$35AU (for an
assumed distance of 140~pc). Its age is estimated to be 1.5~Myrs
\citep{White_1999}. GG\,Tau\,A is surrounded by a circumbinary disk,
which has been extensively studied by means of its visible or mid-infrared
scaterred emission
\citep[e.g.][]{RR_1996,Silber_2000,Krist_2002,McCabe_2002,Krist_2005}
and its direct thermal dust emission in the millimeter domain
\citep[e.g.][]{DGS_1994,GDS_1999}. The higher angular resolution data
clearly illustrate that the inner $\sim$150~AU region of the disk has been
emptied by tidal interactions with the two stars, making GG\,Tau
one of the text-book example of circumbinary disk in young stellar
systems.

\citet{GDS_1999} obtained kinematical information from $^{13}$CO and
HCO$^+$ lines, demonstrating that the disk is in Keplerian rotation (the
redshifted emission being on the western side) around a $\sim$1.3$\msun$
system. These IRAM Plateau de Bure observations also indicated that
the circumbinary disk is composed of a ring with sharp edges
($\sim$70\% of the mass), surrounded by a more extended, colder
disk. The total ring+disk mass is $\sim$0.12$\msun$. Interestingly,
\citet{GDS_1999} noted the continuum 1.4~mm emission also
revealed the presence of material {\it within} the ring, suggesting that there
 are circumstellar disks around the individual sources.
 
In this paper, we present new observations of the millimeter
continuum emission of the GG\,Tau circumbinary disk, which provides a
significant improvement in angular resolution and sensitivity compared
to the \citet{GDS_1999} data and places new constraints on the
properties of the circumbinary and primary disks.
 
\FigVisi{} 

%
%---------------------------------------------------------------------------
%
\section{Observations}

During the winter 2007/2008, new local oscillator components for the
1~mm receiver band were installed on the Plateau de Bure
Interferometer, enabling observations up to a frequency of 270 GHz
(instead of 250~GHz previously). As part of the science verification
program of this upgrade, GG~Tau was observed in January--February
2008. The receivers were tuned in single side band at 267.2~GHz. Only one
polarization channel was used, with a correlator setup covering a
bandwidth of 2~GHz to enable the simultaneous observation of the
HCN(3--2) and HCO$^+$(3--2) lines (which data will not be discussed in
this article).

The source was observed with five antennas in the C configuration
(baseline length from 25 to 175~m) and with six antennas in the A
configuration (baseline length from 130 to 760~m). Both observations
were done in track-sharing mode with two other sources, resulting in a
six-antenna-equivalent total on-source integration time of 4 hours. The
atmospheric conditions were excellent, with a water vapour content of
0.5--1~mm. This translated into a $\sim$200~K system temperature
around transit.

The data processing was performed with the GILDAS\footnote{See
\texttt{http://www.iram.fr/IRAMFR/GILDAS} for more information about
GILDAS.} software. Standard phase and amplitude calibration was
applied to the data, using regular observations of the B0548$+$134
quasar. The atmospheric phase noise was 15--40$^\circ$ for the C
configuration and 30--50$^\circ$ for the A configuration. The absolute
flux calibration was performed using observations of MWC\,349. Since
this project was among the first Plateau de Bure observations at that
high frequency, the flux density of MWC\,349 was known only with a
limited accuracy. We used an extrapolation of the measured flux
densities at lower frequencies.  We carefully checked the relative
calibration of the two tracks by comparing the GG\,Tau flux density on
the overlapping baselines. We estimate the final accuracy of the
absolute flux density to be $\sim20\%$.

The data were imaged using robust weighting, resulting in one of the
highest angular resolutions achieved with the Plateau de Bure
Interferometer, $0.45\times0.25''$ (at $\mathrm{PA}\,19^{\circ}$).  At
the distance of 140~pc, this corresponds to a linear resolution of
$65\times35$~AU.

\TableRing{} 

%
%---------------------------------------------------------------------------
%
\section{Results}

The image reveals a very clear ring structure as well as an unresolved
source located close to the ring center. We performed a fit of a
geometrical ring of uniform brightness to the data, whose results are
given in Table~\ref{tab:ring}. They indicate a narrower ring than
previous studies \citep{GDS_1999}, because of the increase in
angular resolution. Fig.~\ref{fig:visib} displays the real part of the
circularly averaged visibilities as a function of the baseline length
together with the visibility curve of the fitted ring, showing the
excellent agreement with the data. A marginally tighter fit can be
obtained by fitting a ring and a point source (Fig.~\ref{fig:visib},
dashed curve).
%{\bf GIVE RMS}

In addition to standard imaging, we used a more sophisticated
technique to optimize the imaging process for this dataset. 
The dirty beam of the observations had somewhat high sidelobes (at about 25\%
level), owing to the poor $uv$ coverage of the compact C configuration.
As a consequence, the emission within the (bright) ring was difficult
to deconvolve properly. To overcome these difficulties, we subtracted
from the visibilities the fitted geometrical ring
(Table~\ref{tab:ring}) and imaged and deconvolved separately the ring
and the residual data. We then added the two deconvolved maps to
obtain the final image presented in Fig.~\ref{fig:maps}. The advantage
of this method is the level of sidelobes caused by the bright ring 
structure is reduced and the dynamical range of the image is thus increased,
especially in the inner part of the ring. On the final image, the peak
brightness temperature is $\sim$4~K, while the rms noise of the map
is 1.6~mJy/beam (0.24~K), i.e.\ the dynamical range is $>15$.

\FigMaps{}

%
%---------------------------------------------------------------------------
%
\section{Data analysis}

\subsection{The ring}

To analyse the continuum emission of the GG Tau circumbinary ring, we
fitted a {\em physical} model of a ring (in contrast to the {\em
geometrical} model mentioned in the previous section). From a
parametric density and temperature description, the model solves for
the radiative tranfer to predict the observed emission, and performs
a global $\chi^2$ minimization directly in the $uv$ plane \citep[see
e.g.][for a full description of the method]{PDG_2007}. Here, the ring
is assumed to be circular with a constant width and density. Its edges
are modelled as Gaussians decreasing from the ring plateau. The model
also includes a standard disk surrounding the ring, but its
contribution is found to be negligible in reproducing these continuum
data. \citet{GDS_1999} showed that $\sim30\%$ of the mass
lies in the outer disk, but our data lack short baseline measurements
and are therefore certainly filtering out the disk extended emission.

Our best-fit results are presented in Table~\ref{tab:model} and the
corresponding image is displayed in Fig.~\ref{fig:maps}. We could not
constrain the kinetic temperature because the emission is optically
thin within the ring and we therefore assumed a constant temperature
of 35~K, as found from the optically thick \tco~\jdu~ emission by
\citet{GDS_1999}. The derived inclination and position angle are in
perfect agreement with those measured by \citet{GDS_1999}. Compared to
the results of these authors, the ring width is narrower (50~AU
instead of 80~AU) and consequently the surface density has to be higher (2.2
instead of $1.4\,10^{24}$~cm$^{-2}$) in order to reproduce the
observed flux density. The total mass derived from the model is
0.09$\msun$.

The ring width we measure remains only marginally resolved at the
angular resolution of our observations, while the ring edges have a
very narrow width (Gaussians $1/e$ widths $<$10~AU), indicating a very
sharp decrease in the density at the ring edges. Hence, the picture
emerging is that of a very thin ring with a very pronounced
overdensity compared to the inner and outer regions. 

\TableModel{}

\subsection{The central source}

While \citet{GDS_1999} noted the presence of emission within
the GG\,Tau ring, the data presented in this paper clearly separate
for the first time the ring from the central emission.  The central
source that we detect is unresolved at the resolution of these observations
($\sim$50~AU). It is offset by $0.1'' \pm 0.1''$ in R.A. and
$-0.1'' \pm 0.1''$ in declination from the ring center position and
has a flux density of 10~mJy, i.e.\ a brightness temperature of
1.4~K in our synthetized beam.

Using orbital parameters derived by \citet{BD_2005} for the inner
binary, we extrapolated the position of the stars at the time of the
observations, i.e.\ beginning of 2008. These positions (0.25$''$
separation, $-$19$^\circ$ separation angle) are indicated in the top
right panel of Fig.~\ref{fig:maps}. The absolute astrometry of the
stars not being known with enough precision, we forced the center of
mass of the binary \citep[the two stars have similar masses,
0.78$\msun$ and 0.68$\msun$,][]{White_1999} to coincide with the
geometrical center of the ring (cross in Fig.~\ref{fig:maps}). This
hypothesis would not be correct if the ring were elliptical: the center
of mass would then be at one of the ellipse focii. The data and models
do however indicate the ring is circular at our angular
resolution. With these assumptions, the inner continuum emission is
clearly associated with the binary stars. It is thus tempting to
identify this source with the thermal dust emission of one or two
circumstellar disks around the individual stars (an alternative,
although less probable origin would be free-free emission from an
ionized jet).

The continuum emission seems to peak at GG\,Tau\,Aa, although
the formal fitted position indicates a marginal result (2.5
$\sigma$). Both stars exhibit optical emission lines that are typical
of classical T\,Tauri stars and trace accretion processes
\citep{White_1999}, hence indicate that both sources should have an
accretion disk. Our observations suggest that the millimeter continuum
emission of the GG\,Tau\,Ab disk is weaker than that of the Aa disk,
by at least a factor of 3, which points towards a less massive and/or
colder disk. This is consistent with GG\,Tau\,Ab displaying weaker emission 
lines than Aa \citep{White_1999}, which is indicative of less intense accretion activity.

A disk around GG\,Tau\,Aa would be truncated by the relative motion of
Ab to a size typically one third of the orbital distance, i.e.\
$\sim$10~AU. Correcting for beam dilution, the brightness temperature
can thus be $\sim$30~K, which is a lower limit to the disk kinetic
temperature $T_{\rm kin}$. The corresponding mass, assuming optically
thin emission, would be $1.5\,10^{-3}\msun$ (scaling with 30~K/$T_{\rm
kin}$).

\subsection{Streamer?}

The image of GG\,Tau presented in this paper also reveals a continuous
structure connecting the northern part of the circumbinary disk with
the central source. It is very suggestive of a streamer feeding matter
onto the central individual disk(s), as suggested by a number of
theoretical models of circumbinary disk evolution
\citep[e.g.][]{AL_1996}. We note however that this structure is only
marginally detected in our observation (contours are 2$\sigma$ in
Fig.~\ref{fig:maps}) so that it is difficult to draw firm conclusions about 
its existence.

Fig.~\ref{fig:maps} also shows several other low-level (one contour)
extensions of the ring. The reality of these structures remains to be
confirmed, since they are close to the noise level and may also
correspond to deconvolution artefacts in the imaging process. We note,
however, that the position of the feature at the external edge of the 
south-east portion of the ring coincides with the 'kink' seen in
NIR scattered emission by \citet{Silber_2000} or \citet{McCabe_2002}.
Similarly, the western-most extension is seen with several contours,
and may trace a local enhancement of the column density, which could
explain the western part of the ring being brighter at lower frequency
(3.4 mm, \citealt{GDS_1999} or 2.7~mm, \citealt{DGS_1994}).

Finally, the apparent clumpy structure of the ring is not significant:
all peaks correspond to a $<2\sigma$ difference with a uniform ring, as
for instance in the model shown in the right part of the figure.

%
%---------------------------------------------------------------------------
%

\section{Summary}

We have presented a map of the 267~GHz continuum emission in GG\,Tau\,A,
at the unprecedented angular resolution of $0.45\times0.25''$, i.e.\
$65\times35$~AU at the distance of Taurus. These observations confirm
the previous findings of Guilloteau et al.\ (1999) but allow us to place
more accurate constrains on the properties of this system. Our main
conclusions are:

\begin{itemize}

\item The circumbinary dust emission is confined to a circular ring of
width $<$50~AU, with very prononced sharp edges. These observations place
very tight constraints on theoretical model of the
dynamical processes leading to the formation of rings in circumbinary
systems \citep[e.g][]{BD_2005,BD_2006}. GG\,Tau is 1.5~Myrs old, but,
as noted by \citet{BD_2006}, the observed ring may be a
transient feature that will undergo significant evolution in the
next Myrs.

\item The data very clearly diffentiate the circumbinary disk from the
emission located within the disk: an unresolved source is detected at
the position of GG\,Tau\,Aa. This could be the first direct detection
of of a circumstellar disk surrounding one of the stars of the central
binary system. The mass of the circumprimary disk(s) is
$1.5\,10^{-3}\,\msun$, about 80 times less than the mass of the
circumbinary material.

\end{itemize}

These observations confirm that GG\,Tau\,A is one of the best known
and most interesting examples of a disk surrounding a young binary
T\,Tauri system. It will be a prime target for future high angular
studies with an instrument like ALMA.

%
%---------------------------------------------------------------------------
%
\begin{acknowledgements}
We acknowledge help from the Plateau de Bure staff for performing the
observations.
\end{acknowledgements}

%
%---------------------------------------------------------------------------
%

\bibliography{15682}
\bibliographystyle{aa}
\end{document}